\documentclass[letterpaper,conference]{IEEEtran}
\ifCLASSINFOpdf
\else
\fi
\PassOptionsToPackage{bookmarks=false}{hyperref}
\usepackage[draft]{hyperref}
\usepackage{algorithmic}
\usepackage{setspace}
\usepackage{color}
\usepackage{xcolor}
\usepackage{enumerate}
\usepackage{srcltx}  
\usepackage{amsmath}
\usepackage{amsthm}
\usepackage{amssymb}
\usepackage{graphics}
\usepackage{cite}
\usepackage{psfrag}
\usepackage{epsfig}
\usepackage{subfigure}
\usepackage{mdframed}
\usepackage{float}
\PassOptionsToPackage{bookmarks=false}{hyperref}
\usepackage[draft]{hyperref}

\theoremstyle{definition}

\newcommand{\no}{\nonumber}

\usepackage{algorithm}
\usepackage{algorithmic}

\makeatletter
\newcommand\fs@norules{\def\@fs@cfont{\bfseries}\let\@fs@capt\floatc@ruled
  \def\@fs@pre{}%
  \def\@fs@post{}%
  \def\@fs@mid{\kern3pt}%
  \let\@fs@iftopcapt\iftrue}
\makeatother
\floatstyle{norules}
\restylefloat{algorithm}

\begin{document}
%
\title{The Effect of Interference in Vehicular Communications on Safety Factors}
%
%
%

\author{\IEEEauthorblockN{Ali Rakhshan}
\IEEEauthorblockA{School of Electrical and\\Computer Engineering\\
Johns Hopkins University\\
Baltimore, Maryland\\
Emails: arakhsh1@jhu.edu}}
\maketitle

\begin{abstract}
This paper studies how the interference of vehicular communications affects the safety of vehicles in a vehicular ad hoc network. Different signal propagation models with and without carrier sensing are considered for the dissemination of periodic safety messages. Then, by employing the results for different packet success probability of the vehicles, we compare the expected collision probability of vehicles for different signal fading models. Our results show how the collision probabilities of vehicles in the network vary with respect to different models and designs.
\end{abstract}

\begin{IEEEkeywords}
Vehicular Networks, Wireless Networks, Ad Hoc Networks, Interference, Safety, Carrier Sensing.
\end{IEEEkeywords}

%
\IEEEpeerreviewmaketitle

\section{Introduction}

Despite the fact that new technologies are incorporated into transportation systems every day, the improvement of drivers' safety is not proportional to the rapid rise of advanced technology.
Statistics from the National Highway Traffic Safety Administration (NHTSA) in 2013 \cite{National-a} report over five million crashes in the U.S., causing over two million injuries and more than 30,000 fatalities.

 NHTSA research proposes a way to improve the effectiveness of collision warning systems by employing Vehicular Ad Hoc Networks (VANETs)\cite{National-b}. VANETs provide on-board units with  two types of vehicular communications. These include the communication between vehicles in close proximity to each other as well as communication between the infrastructure on the road and the vehicle.
 The Federal Communications Commission has allocated 75 MHz of spectrum in the 5.9 GHz band for Dedicated Short Range Communications (DSRC).
 To serve as the groundwork for these communications, the IEEE 802.11p standard was published in the year 2010 \cite{802.11p} for Wireless Access in Vehicular Environments (WAVE).

There are different types of collisions that potentially happen on roadways. Of these, rear-end collisions represent 28\% of the crashes \cite{National-b}. This type of collision symbolizes the most important line-of-sight environment for vehicular communications. Hence, the majority of crashes actually occur in non-line-of-sight environments.

A number of models exist to describe the statistics of the amplitude and the phase of multi-path fading signals. The Nakagami-m distribution has some advantages over other models like Rayleigh fading and Rician fading.
However, many papers have considered the simpler models to analyze the interference at the expense of losing the required accuracy for drivers' safety analysis. Carrier sensing has also been a neglected factor in the safety packets delivery analysis.

Our main contributions in this paper are as follows:
\begin{itemize}
 \item We analytically study the success probability of safety packets by taking the multi-user interference, path loss, and two different types of signal fading models into account.
 \item We also consider the scheme in which each node senses the channel at the beginning of each slot.
 \item We compare the vehicle collision probabilities of the network for various discussed models.
\end{itemize}

The remainder of this paper is organized as follows. Section \ref{section:Background} reviews the past literature in the interference modeling of VANET. In section \ref{section:assumptions} and section \ref{section:analysis}, the assumptions and the analysis to characterize the interference are presented. Finally, the numerical results are illustrated in section \ref{section:result}.

\section{Background And Literature Review}\label{section:Background}

Our goal is to examine how many transmissions on average are required for a vehicle to receive the desired safety packet. Chang et al. defined a new metric named the probability of reception failure (PRF) for a scheme in which nodes transmit with a given probability in each slot. However, they neglected fading and only considered the strongest
interferer in their analysis \cite{Ref:Chang}. Garcia-Costa et al. proposed a stochastic model in which they obtained the average number of collisions in a chain of vehicles.
Each vehicle was assumed to be equipped with a collision warning system; however, the distribution of the
safety packets delivery was fixed and unrealistic for every MAC scheme \cite{Ref:Garcia-Costa}. Carbaugh et al. also studied the rear-end collisions of automated and manual highway systems.
Yet, they assumed a fixed communications delay (fixed packet success probability) of 300, 150, and 120 milliseconds for autonomous, low-cooperative, and high-cooperative vehicles, respectively, an assumption which might not be realistic  \cite{Ref:Carbaugh}.
Haas et al. simulated two vehicular safety applications and determined the effect of various communication parameters on vehicle crash avoidance through traffic simulations\cite{Ref:Haas}. However, they did not develop any mathematical framework for the interference modelling of VANET.
Finally, we studied the concept of different channel access for the vehicles in a highway scenario, but we only analyzed a specific scenario \cite{Rakhshan:CISS2, Rakhshan:TWC, Rakhshan:Infocom, Rakhshan:CISS}.

There are major differences between this paper and others. First, most of the studies which examine different interference models are only simulation-based (e.g. \cite{Ref:Islam} and \cite{Ref:Killat}). However, we want to find insights through the analysis on how different parameters can actually change the delivery of packets and thus the vehicle collision probability. Clearly, the results obtained from both different simulators and analysis are only an approximation of reality. Second, we will demonstrate the effect of carrier sensing (or non-independent channel access of vehicles) on the packet success probability which is usually neglected in the analysis. Third, the channel access is assumed to be equal for different vehicles in the analysis. Although this assumption seems realistic based on the current vehicles equipped with DSRC antennas, in the near future this assumption may need to be relaxed. In other words, the channel access of different drivers may depend on the safety of their vehicles in future designs. Hence, we assume the vehicles can transmit at different rates.

\section{Assumptions}\label{section:assumptions}
Communications between vehicles can help drivers react properly to deceleration events, especially when a driver cannot either observe or perceive the deceleration of other vehicles due to low visibility, high unexpectedness of the incident, defected brake lights, or many of the other distractions that nowadays exist on the roads. However,
we need to know the communication interference of other vehicles' signals in order to find any other important safety factors in our design, such as packet delivery success probability and vehicle collision probability.
It has been shown that the Nakagami fading model describes the interference more accurately than other well-known relatively simple models for vehicular ad hoc scenarios \cite{Ref:Islam},\cite{Ref:Killat}. Other papers examine this interference only through simulations, but we study it analytically to find critical factors in drivers' safety. 

It is noteworthy that due to both the short length of packets
and the broadcast nature of communications, the 4-way handshake anticipated by the standard is not efficient
for the dissemination of periodic safety messages.
RTS/CTS and ACK message
exchanges worsen the hidden node problem thus leading to
higher probability of packet collisions \cite{Ref:Farnoud}.
Furthermore, only the protocols which do not need a detailed description of the network topology to
schedule packet transmissions are effective because the topology of VANETs is immensely dynamic. Repetition-based protocols not
only reveal this property, but also fight packet collisions due
to the problem of hidden nodes. A similar approach has been used in other papers, e.g. in \cite{Ref:Xu} and \cite{Ref:Farnoud}.

In the next section, first we employ repetition-based protocols for the dissemination of periodic safety messages. Second, we consider the Slotted Asynchronous P-persistent with carrier sensing (SAP/CS) scheme. Hereafter, we do not restrict our analysis to any specific geometry unless explicitly stated otherwise.

\section{Analysis}\label{section:analysis}
We need to know the communication interference of other vehicles' signals in order to find any other important safety factors in our design, such as packet delivery success probability and vehicle collision probability.

Path loss and Nakagami-m fading are taken into account for formalizing the signal propagation characteristics. If the nodes transmit with unit power, the received power at distance $r$ is $hr^{-\alpha}$ where $\alpha(>1)$ is the path loss exponent and $h$ is the fading coefficient. We assume that the magnitude of the signal that has passed through the transmission medium will vary randomly according to the Nakagami-m distribution. This is a valid assumption because the sum of multiple independent and identically distributed (i.i.d.) Rayleigh-fading signals, which have a Nakagami distributed signal amplitude, have been shown to be an efficient interference model for multiple sources \cite{Nakagami}. Since the amplitude of the received signal is a Nakagami-m distributed random variable, $h$ has gamma distribution with mean $\lambda$:
\begin{align}
\no f_H(h)=\frac{1}{\Gamma(m)}\left(\frac{m}{\lambda}\right)^m h^{m-1}e^{\frac{-mh}{\lambda}} \quad h\geq 0
\end{align}
where $\Gamma(m)$ is the gamma function for integer shape factor $m$. Assuming that a vehicle transmits a packet, the per-hop transmission success probability can be calculated as follows ($E(h_i)=\lambda=1$):
\begin{align}\label{eq1:nakagami_main}
 P_S&=\mathbb{P}\left(\frac{S}{I}>\beta\right)
\end{align}
\begin{align}\label{eq2:nakagami_main}
\no P_S&=\mathbb{P}\left(\frac{hr^{-\alpha}}{\sum_{i=1}^{n}b_ih_ir_i^{-\alpha}}>\beta\right)\\
\no &=\int \mathbb{P}\left(h>\beta r^{\alpha}I|I=i\right)f_I(i) di\\
 &=E_I \left[1-\frac{1}{\Gamma(m)}\gamma(m,m\beta r^{\alpha}I)\right]
\end{align}
 \begin{align}\label{eq3:nakagami_main}
 &=1-\frac{1}{\Gamma(m)}m^m\sum_{k=0}^{\infty} \frac{(-m)^k\beta^{k+m}}{k!(k+m)} E[r^\alpha I]^{(k+m)}
 \end{align}
 \begin{align}
 \no &=1-\frac{1}{\Gamma(m)}\sum_{k=0}^{\infty}\frac{(-1)^k\beta^k} {k!(m+k)[(m-1)!]^k} \cdot \prod_{i=1}^n p_i
 \end{align}
 \begin{align}\label{eq4:nakagami_main}
 \no &\cdot\sum_{k_1+k_2+\cdots+k_n=k} \left(\begin{array}{c} k \\ k_1,k_2,\cdots,k_n \end{array}\right)(m+k_i-1)! \\
 &\cdot E\left[\left(\prod_{j=1}^n \left(\frac{r}{r_j}\right)^{\alpha k_i}\right)\right].
\end{align}
\begin{table}[t]
\caption{Definitions of the variables in Equations \ref{eq1:nakagami_main}, \ref{eq2:nakagami_main}, \ref{eq3:nakagami_main}, \ref{eq4:nakagami_main}}
\label{table_def}
\begin{center}
\begin{tabular}{|c|c|}
  \hline
   $S$ & Desired signal power\\
  \hline
   $I$ & Interference power at the receiver\\
   \hline
   $\alpha$ & Path loss exponent\\
  \hline
  $\beta$ & SIR decoding threshold\\
  \hline
  $p_i$  & Transmission probability of node $i$\\
  \hline
  $b_i$  & Bernoulli random variable with probability $p_i$\\
  \hline
   $r_i$  & Distance from the interferer $i$ to the receiver\\
  \hline
  $r$  & Distance between the transmitter and the receiver\\
  \hline

  $n$ & Number of vehicles\\

  \hline

  $h_i$ & Fading coefficient of interferer $i$\\

  \hline

\end{tabular}
\end{center}
\end{table}
The definitions of the variables are given in Table \ref{table_def}.
 A fixed coding scheme is considered in Equation \ref{eq1:nakagami_main} that requires the SIR at the receiver to be greater than some threshold which is chosen based on IEEE 802.11p tables \cite{802.11p} (e.g. Table \ref{table_Data_Rate}). $S$ denotes the power of the main signal which faces interference from the other vehicles with the accumulative power of $I$. Equation \ref{eq2:nakagami_main} is then obtained by substituting the definitions of the transmitter signal strength and the interference signal strength in Equation \ref{eq1:nakagami_main}. Each of the vehicles is either in the transmitting mode with probability $p_i$ or in the receiving mode with probability $1-p_i$. A Bernoulli random variable, $b_i$, represents the state of the transmitter vehicle $i$.
 Equation \ref{eq3:nakagami_main} is resulted by employing the following convergent series of the incomplete gamma function to cancel $h$:
\begin{equation}
\no \gamma\left(m,m\beta r^{\alpha}I\right)=\frac{1}{\Gamma(m)}\sum_{k=0}^{\infty}\frac{\left(-m\beta r^{\alpha}I\right)^k} {k!(m+k)}.
\end{equation}
Finally, the multinomial expansion and characteristic functions of fading random variables leads us to Equation \ref{eq4:nakagami_main}.

The obtained packet success probability equation clearly holds for all the possible geometries. For $m=1$, Equation \ref{eq3:nakagami_main} will be equal to:
\begin{align}\label{eq:Rayleigh}
P_S&=\prod_{i=1}^n\left[1-p_i+\frac{p_i}{1+\beta \left(\frac{r}{r_i}\right)^\alpha}\right].
\end{align}
which is the packet success probability equation when the Rayleigh fading model is employed \cite{Rakhshan:CISS2}.
If the time slots in which nodes transmit are not synchronized, this scheme is named \textit{Slotted Asynchronous P-persistent} (SAP).
In this case, an interferer can potentially interfere with \textit{at most} two time slots of another transmission. Hence, the transmission probability for the interferers is:
\begin{equation}
 p^\prime_i=p_i+p_i-p_i \cdot p_i \simeq 2p_i. \label{Eq:async}
\end{equation}
Since the probabilities are small, this approximation is close to the real value.

Up to this point, we have assumed that each vehicle transmits independently of all other vehicles. However, in order to reduce the probability of packet collisions, we study a scheme in which each vehicle transmits only if it finds the channel idle (sensing). Our goal is to find the packet success probability under the Nakagami-m fading model by employing the SAP/CS scheme.
\begin{figure}[!t]
\centering
\includegraphics[width=3.5in]{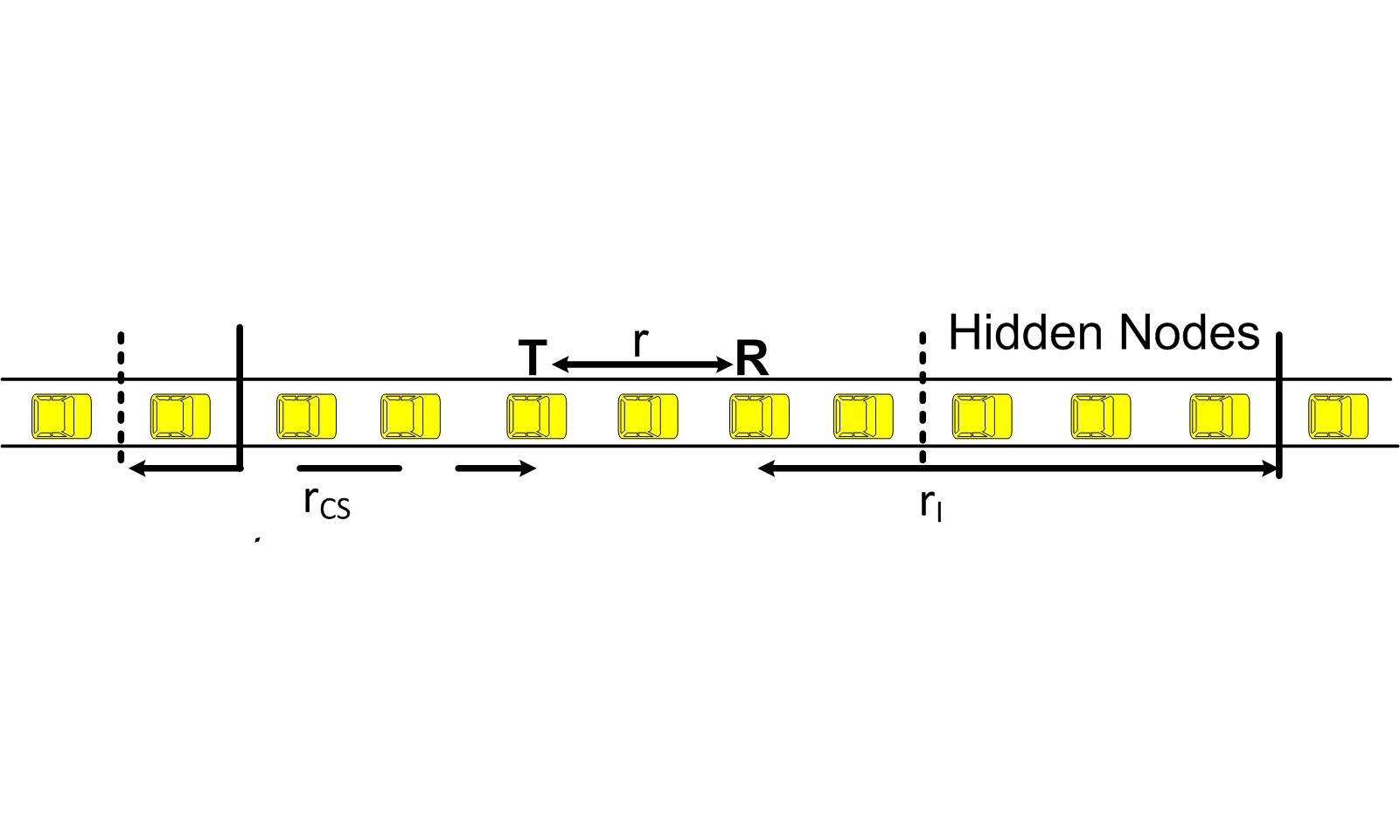}
\caption{A chain of vehicles which employ SAP/CS MAC scheme. T, R, $r_{CS}$, $r_I$ represent the transmitter vehicle, the receiver vehicle, the carrier sensing distance, and the distance in which the vehicles can cause interference at the receiver vehicle, respectively.} 
\label{fig:CS}
\end{figure}
To make the analysis feasible, we start with:
\begin{equation}
\no P_s=P_t \cdot P_{s|t}.
\end{equation}
$P_t$ represents the probability that node $T$ accesses the channel, i.e. finds it idle and transmits. $P_{s|t}$ is the packet success probability at vehicle $R$, given that vehicle $T$ accesses the channel (Fig. \ref{fig:CS}). We define the carrier sensing distance as $r_{CS}$. A vehicle can transmit if and only if no other vehicle transmits within $r_{CS}$ distance of it. The number of vehicles within this radius is called $n_{CS}$:
\begin{equation}
 P_t \approx p_{T}\prod_{i=1}^{n_{CS}}(1-p_i).\label{Eq:carrier}
\end{equation}
in which $p_{T}$ represents the channel access probability of the transmitter vehicle.
The approximation is sufficiently tight because transmission probabilities are small despite the transmissions not being independent. If the probabilities are not small, Equation \ref{Eq:carrier} denotes an upperbound for $P_t$.

Next, we need to find the radius of a disk centered at $R$ in which any active node can cause interference at $R$. According to the SIR-based reception model, there must be $\frac{hr^{-\alpha}}{h_ir_i^{-\alpha}} > \beta$ where $h$ and $h_i$ are the respective Nakagami-m fading components of the interference model, and $r$ and $r_i$ are the distance between the transmitter and the receiver and the distance between the interferer $i$ and the receiver. Therefore, we have:
\begin{align}
\no r_I \approx r\beta^{\frac{1}{\alpha}}\mathbb{E}\left[h^\frac{-1}{\alpha}\right]\mathbb{E}\left[h_i^\frac{1}{\alpha}\right].
\end{align}
By employing the concept of fractional moments, we obtain:

\begin{align}
\no r_I &\approx r\cdot \beta^{\frac{1}{\alpha}} \frac{\Gamma(m+\frac{1}{\alpha})}{\Gamma(m)} \frac{\Gamma(m-\frac{1}{\alpha})}{\Gamma(m)}.
\end{align}
\begin{align}
\no &= r\cdot \beta^{\frac{1}{\alpha}} \frac{\frac{\pi}{\alpha}}{\Gamma^2(m)}\csc(\frac{\pi}{\alpha}).
\end{align}
For the Rayleigh fading scenario,
\begin{align}
\no r_I &\approx r\cdot \beta^{\frac{1}{\alpha}} \frac{\pi}{\alpha}\csc(\frac{\pi}{\alpha}).
\end{align}
In the absence of fading, $r_I \approx r\beta^{\frac{1}{\alpha}}$. For usual values of $\alpha$, the $r_I$ is greater than when there is no fading. When vehicle $T$ transmits, only the hidden nodes whose activities are not sensed by node $T$ can cause outage at node $R$ (see Fig. \ref{fig:CS}). If there are $x$ hidden nodes and $N_i$ represents the event that the $i$th hidden node does not transmit, then $P_{s|t}$ is equal to
\begin{align}
\no P_{s|t}&=\mathbb{P}\left(\bigcap_{i=1}^x N_i\right)\\
\no        &=1-\mathbb{P}\left(\bigcup_{i=1}^x N^c_i\right)\\
           &=1-\sum_{i=1}^x \mathbb{P}(N_i^c). \label{Eq:intersection}
\end{align}
The last equality is true when $N_i^c \bigcap N_j^c = \O$. This condition holds true since the MAC scheme does not allow the hidden nodes to transmit simultaneously for the practical values of $r_{CS}$ and $r_I$. For a one lane case, the packet success probability of the transmitter $T$ at the receiver $R$ can be approximated as:
  \begin{equation}
   P_s \approx \left\{
  \begin{array}{l l}
    p_{T}\prod^{n_{CS}}_{i=1}(1-p_i)\left[1-\sum_{i=1}^{N(r+r_I-r_{CS})}p_i^{'}\right]  &  \quad\\   max\left(r_I-r,\frac{r+r_I}{2}\right)\leq r_{CS}<r+ r_I\\
      &  \quad   \\
    p_{T}\prod_{i=1}^{n_{CS}}(1-p_i) &  \quad\\ r_{CS} \geq r+r_I
  \end{array} \right.\label{Eq:Packet_Success}
  \end{equation}
$N(r+r_I-r_{CS})$ represents the number of hidden nodes in the hidden area $(r+r_I-r_{CS})$. The optimal carrier sensing distance is $r^{*}_{CS} \approx r + r_I$. Here, $r_I-r\leq r_{CS}$ represents the scenario in which there is no hidden node to the left of node $T$ (in Fig. \ref{fig:CS}). In order for Equation \ref{Eq:intersection} to hold, $\frac{r+r_I}{2}$, which is the maximum distance between the hidden nodes, must be less than $r_{CS}$ to force the vehicles not to be active together. If the time slots are not synchronized, the interferers can cause outage over two consecutive time slots with the probability obtained by Equation \ref{Eq:async}.
\section{Numerical Results}\label{section:result}
\begin{table}[t]
\caption{IEEE 802.11P data rates and corresponding SIR decoding thresholds}
\label{table_Data_Rate}
\begin{center}
\begin{tabular}{|c|c|c|c|c|c|c|c|}
  \hline
   $R$ (Mbps) & $3$  & $4.5$ & $6$ & $9$ & $12$ & $18$ & $24$   \\
   \hline
   $\beta$ (db) & $5$ & $6$ & $8$ & $11$ & $15$ & $20$ & $25$   \\
  \hline
\end{tabular}
\end{center}
\end{table}
\begin{table}[t]
\caption{Simulation Parameters. Data rate and SIR decoding threshold are chosen based on \cite{Ref:Xu}}
\label{table_sim_par}
\begin{center}
\begin{tabular}{|c|c|}
  \hline
   Vehicle Distribution & Equal Distance\\
  \hline
   Velocity & $20 \frac{m}{s}$\\
   \hline
   Deceleration rate & $[-6,-9] \frac{m}{s^2}$\\
  \hline
  Total number of selected vehicles  & $25$\\
  \hline
   SIR decoding threshold  & $8$ dB\\
  \hline
   R=Data rate  & $6$ Mbps\\
  \hline

  Number of Obstructive Vehicles  & $4$\\

  \hline

  L=Packet length  & $250$ Bytes\\
  \hline

  Reaction times of drivers  & $ln\mathbf{N}(0.17,0.44)$\\

  \hline
  
  Distance between neighbors & $25m$\\
  
  \hline
\end{tabular}
\end{center}
\end{table}
\begin{figure}[!t]
\centering
\includegraphics[width=3.5in]{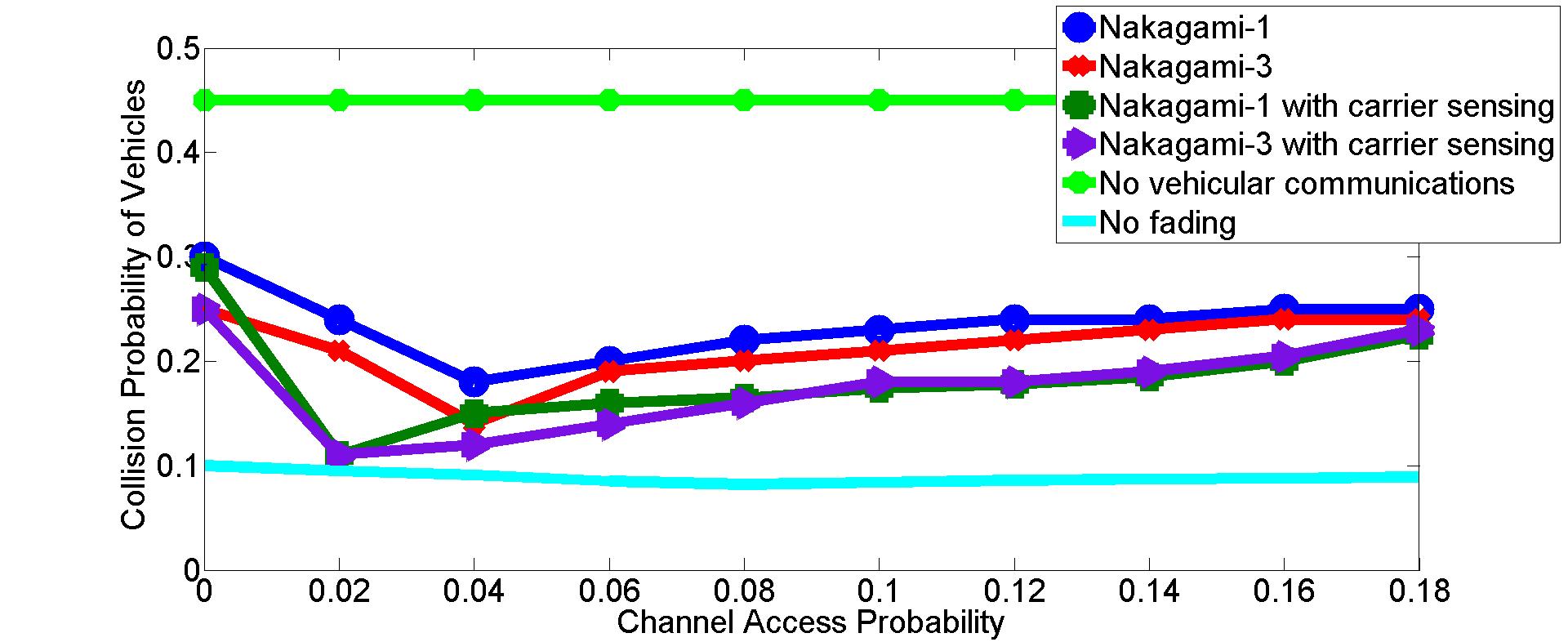}
\caption{The average collision probability of vehicles in the network versus channel access probability.} 
\label{fig:CP}
\end{figure}
In this section, we want to compare the performance of different designs in a highway scenario considering both discussed cases, with and without carrier sensing. Table \ref{table_sim_par} shows all the values assigned to different parameters.
In a chain of vehicles, we assume transmissions across the chain are partially obstructed by some vehicles that are chosen uniformly in our Monte Carlo simulations. In other words, the selected vehicles disrupt the line-of-sight environment for the specific scenario and divide the chain into smaller chains.
The collision probability is calculated based on the equations of motion. The drivers can react to the deceleration of their leading car with reaction time chosen randomly from the lognormal distribution with parameters $\mu= 0.17$ and $\sigma = 0.44$ (see \cite{Koppa:HumanFactors}). The vehicles transmit with equal channel access probability and the distance between neighbor vehicles equals to $25m$ (see \cite{Ref:May}) . Therefore, the packet success probability is obtained by employing Equations \ref{Eq:Packet_Success}, \ref{eq:Rayleigh}, and \ref{eq4:nakagami_main}. Also, each vehicle decelerates as soon as it is informed with a rate chosen uniformly at random from the interval $[-6,-9] \frac{m}{s^2}$.

 We need to compute the time it takes for a message to be received by vehicle $i$ in the part of the chain that doesn't include any obstructive vehicles.
As a result, the successful reception at vehicle $V_i$ has a geometric distribution with parameter $P_s(i) \cdot p_{tr} \cdot (1-p_i)$. Here, $p_{tr}$, $p_i$, and $P_s(i)$ represent the channel access probability of the transmitter, the channel access probability of the desired receiver ($i^{th}$ vehicle), and the packet success probability at the desired receiver, respectively.

Let's assume a chain of vehicles is moving in a certain direction on a highway. We name the first vehicle in the chain $V_0$ and the following vehicles $V_1$, $V_2$, ... respectively.
Clearly, it takes longer time for the vehicles far away from $V_0$ to receive the packets due to delay.
The far vehicles on the highway (for example $V_j$) receive the messages about the deceleration of $V_0$ from the vehicles $V_1 \cdots V_{j-2}$ as well. $V_{j-1}$ is not included since $V_j$ can see the brake lights of $V_{j-1}$ with no need for vehicle-to-vehicle communication. Taking all of the above into account, the average delay of reception at vehicle $V_i$ is:

\begin{align}
\no D(i)&=\min(\min_{(j \in {1,\cdots,i-2})} \frac{L}{R}\frac{1}{P_s(j)p_{0}(1-p_j)}\\
\no &+\tau_j+\frac{L}{R}\frac{1}{P_s(i)p_{j}(1-p_i)}, \frac{L}{R}\frac{1}{P_s(i)p_{0}(1-p_i)},
\end{align}
\begin{align}
 &\frac{L}{R}\frac{1}{P_s(i-1)p_{0}(1-p_{i-1})}+\tau_{i-1})\quad i>2. \label{eq:communication delay}
\end{align}
$R$ represents the data rate which is chosen from TABLE \ref*{table_Data_Rate} while $L$ denotes the packet length. In addition, $\tau_j$ denotes the reaction time of the $j^{th}$ driver in the chain. Each vehicle can retransmit the transmitter's safety packet after its corresponding driver applies the brake. Therefore, the communication delay is actually the minimum of three parts. The first part of Equation \ref{eq:communication delay} symbolizes the retransmission of the safe packets by the middle vehicles while the second part represents the direct communication between the transmitter and the desired receiver. The last part assumes the vehicle in front of the receiver receives the safety packet and the driver of the desired receiver can see the brakes lights of the vehicle.
The allowable number of transmission opportunities within the tolerable delay period is:
\begin{align}
\no D&=\lfloor{\frac{T(i)R}{L}}\rfloor.
\end{align}
 $T(i)$ denotes the maximum tolerable delay to inform vehicle $V_i$ in a chain of vehicles.

The average collision probability of vehicles in the network is illustrated in Fig. \ref{fig:CP}. When the channel access probability is around $0.04$, the communication with carrier sensing is almost the same as the scenario without carrier sensing. 
Also, for large channel access the difference between the schemes with and without carrier sensing shrinks. In these two ranges of channel access, there is less than 10\% improvement in the collision probability. Since Equation \ref{Eq:carrier} represents an upperbound for the success probability, the resulting reduction achieved by employing the carrier sensing is the maximum possible difference between the two curves. Therefore, it confirms that carrier sensing could be noticeably beneficial only in a specific range. Furthermore, the minimum collision probability can be achieved at lower channel access for the vehicles when carrier sensing design is employed compared to  when it is not.  
Fig. \ref{fig:CP} also depicts that employing the more accurate model (Nakagami-3) results in lower collision probability especially when carrier sensing is used.
\section{Conclusion}
In this paper, we studied the effect of the vehicular communications interference on the delivery of safety packets in vehicular ad hoc networks. 
We derived the approximated packet success probability for different scenarios. 
Our results illustrated that the difference between the average collision probability in the network for two scenarios with and without carrier sensing is minimized almost at the minimum point of the latter scenario. The difference reaches local maxima 
when the channel access value increases or decreases compared to the minimum point. 

\end{document}